\documentclass[fleqn,usenatbib]{mnras}

\usepackage{newtxtext,newtxmath}
\usepackage{ulem}

\usepackage[T1]{fontenc}

\DeclareRobustCommand{\VAN}[3]{#2}
\let\VANthebibliography\thebibliography
\def\thebibliography{\DeclareRobustCommand{\VAN}[3]{##3}\VANthebibliography}

\usepackage{graphicx}	% Including figure files
\usepackage{amsmath}	% Advanced maths commands
\title{Optical Variability, Rotation Period and Inclination Angle of the M9.5 dwarf BRI 0021-0214}

\author[S.Dulaimi et al.]{
	S. Dulaimi,$^{1,2,4}$\thanks{E-mail: salam.dulaimi@ced.nahrainuniv.edu.iq }
	A. Golden$^{1,2,5}$
	R. P. Boyle,$^{3}$
	and R. F. Butler$^{1,2}$
	\\
	$^{1}$Centre for Astronomy, University of Galway, University Road, Galway, Ireland\\
	$^{2}$School of Natural Sciences, University of Galway, University Road, Galway, Ireland\\
	$^{3}$Vatican Observatory Research Group, Steward Observatory, University of Arizona, Tucson, AZ 85721, USA\\
	$^{4}$Department of Physics, College of Science, Al-Nahrain University, Jadriya, Baghdad, Iraq \\
	$^{5}$Armagh Observatory and Planetarium, College Hill, Armagh, BT61 9DB, Co. Armagh, Northern Ireland
}

\date{Accepted XXX. Received YYY; in original form ZZZ}

\pubyear{2022}

\begin{document}
	\label{firstpage}
	\pagerange{\pageref{firstpage}--\pageref{lastpage}}
	\maketitle
	
	\begin{abstract}\\
		We report $I$-band photometric observations of the radio-detected M9.5 dwarf BRI 0021-0214, obtained with the Galway Ultra Fast Imager (GUFI) on the 1.8m Vatican Advanced Technology Telescope VATT at Mt. Graham International Observatory, Arizona. In total, 19 hours of observations over a 73 day baseline were obtained. BRI 0021-0214 was shown to exhibit modulated emission with a period of $ 3.052 \pm 0.004$ hours  with a mean amplitude variability of 0.0044 mag. When combined with rotational velocity data obtained from previous work, our newly discovered rotation period gives an inclination angle of 51.7$^{+5.0}_{-4.5}$ degrees for the rotation axis of BRI 0021-0214 relative to our line of sight. Previous studies have reported that the most plausible cause for optical variability from this dwarf is a consequence of suspended co-rotating dust clouds in its atmosphere. However reports of enhanced H$_{\alpha}$ and intermittent coherent radio emission suggest the possibility of auroral activity in its magnetosphere. Further, more coordinated multiwavlength observations of this dwarf could fully resolve the nature of this elusive  rapid-rotator object's observational properties.

	\end{abstract}
	
	%%%%%%%%%%%%%%%%%%%%%%%%%%%%%%%%%%%%%%%%%%%%%%%%%%
	
	%%%%%%%%%%%%%%%%% BODY OF PAPER %%%%%%%%%%%%%%%%%%
	
	\section{Introduction}
	
	%Very low mass stars and brown dwarfs (spectral types  $\sim$ M7 and later) - also known as ultracool dwarfs - congregate at the tail end of the Hertzsprung\text{-}Russell (H\text{-}R) diagram \citep{kirkpatrick1997ultra}. 
	
	Many photometric surveys \citet{tinney1999searching,martin2001photometric,bailer2002dust,gelino2002dwarf,rockenfeller2006multiband,littlefair2008optical, mclean2011periodic} have demonstrated the presence of modulations in both optical and near infrared lightcurves of objects that span the substellar boundary beyond spectral type M9. These variations  are notable by either a consistently periodic, albeit weak, signal with amplitude $\leq$ 1.5 \% \citep{artigau2009photometric, radigan2012large, gillon2013fast, crossfield2014global} or a more aperiodic signature with variable modulation  \citep{buenzli2014brown,  wilson2014brown, radigan2014independent, metchev2015weather}, on timescales ranging from 1.4 - 20 hours.
	Beyond the M/L transition, dust grain formation is expected to occur in the  photospheres of ultracool dwarfs \citep{tsuji1996dust,allard2001limiting} and so is likely to  play a role in any observed optical variability for cooler objects.\\

	H$_{\alpha}$ and X-ray observations have also revealed that the majority of late M to early L ultracool dwarfs are still capable of producing and sustaining significant magnetic field structures \citep{hawley1996palomar,berger2001discovery, berger2002flaring,berger2005magnetic,burgasser2005quiescent, osten2006radio,hallinan2006rotational,berger2006radio, schmidt2007activity,west2011sloan,schmidt2015boss} suggesting that magnetic activity may also contribute to the optical variability observed \citep{hooten1990photometry, henry1995automated, strassmeier125rodono,rockenfeller2006multiband,lane2007rotational}. Several studies have shown  that the detected optical variability in several ultracool dwarfs can be best explained a combination of both magnetic cool spots and atmospheric dust clouds \citep{apai2013hst, heinze2013weather, gizis2015kepler}. Indeed, work in particular by \citep{miles2015linear} demonstrated that optical linear polarization variability is correlated with the radio emission (and so magnetic field) derived rotation period of the M8.5 dwarf TVLM 513-46546, though the likely source remains unresolved.\\

	\begin{table} 
		\centering
		\begin{tabular}[c]{ccccccccc}
			
			\hline
			Parameter & BRI 0021-0214 & References \\
			\hline
			Spectral type & M9.5 &\citet{leggett2001infrared} \\
			Distance (pc) & $\sim$11.55  & \citet{reid1999bri} \\
			$\it T_{eff}$ (K) & $2385 \pm 77$ & \citet{gonzales2018understanding}\\
			log ($\it L_{bol}$/ $\it L_{\odot}$) (dex) & \text{-} $3.44 \pm 0.04$ & \citet{gonzales2018understanding} \\
			{\it v sin i}  (\text{km} $\text{s}^{-1}$)  &  $34.2 \pm 1.6$  &  \citet{crossfield2014doppler}\\
			Radius ($\it R $/$\it R_{\odot}$) &  $0.109  \pm 0.005 $ &  \citet{filippazzo2015fundamental}\\
			Estimated Mass ($\it M_{\odot}$) &  $0.075   \pm 0.010 $ &  \citet{filippazzo2015fundamental}\\
			Inclination Angle  ($^\circ$) & 51.7$^{+5.0}_{-4.5}$ & This work\\
			Rotational Velocity (\text{km} $\text{s}^{-1}$) & $43.6 \pm 2$ & This work\\
			% Ray: final line (Rotational Velocity) added, per referee request
			\hline
			\centering
		\end{tabular}
		\caption{\label{tab:BRI properties}A summary of the physical properties of BRI 0021-0214.}
		
	\end{table}

	A subset from M8 $\rightarrow$ T6, all typically fast rotators with rotational periods of order  of hours, have been shown to possess powerful, stable and highly structured kG magnetic fields as confirmed by the detection of powerful coherent beams of radio emission at cm wavelengths caused by electron cyclotron maser (ECM) processes associated with aurora in their magnetospheres \citep{hallinan2015magnetospherically, kao2016auroral, pineda2017panchromatic}. The associated auroral currents impact the dwarf photosphere/atmosphere, resulting in localized H$_{\alpha}$\ \& optical/NIR continuum emission. The co-rotation of the auroral regions within the stable magnetic structures of the dwarf magnetosphere then ensures an observable modulated signal in both radio and optical bands.
	Whilst there is a widespread acceptance of the presence of significant magnetic activity associated with fully convective substellar objects, with non-axisymmetric fields of order $\sim$ 10$^{1-2}$ G complemented by axisymmetric poloidal fields of order $\sim$ 10$^{3}$ G \citep{kochukhov2021}, their location and (particularly in the case of the dominant dipolar fields) their orientation with respect to the dwarf's rotational axis is unclear. By convention, co-alignment is presumed; however there are incidences where this may not be the case \citep{williams2017}.\\

	\begin{figure*}
		\centering
		%\resizebox{{\columnwidth}{!}{
		\includegraphics[width=18.2cm]{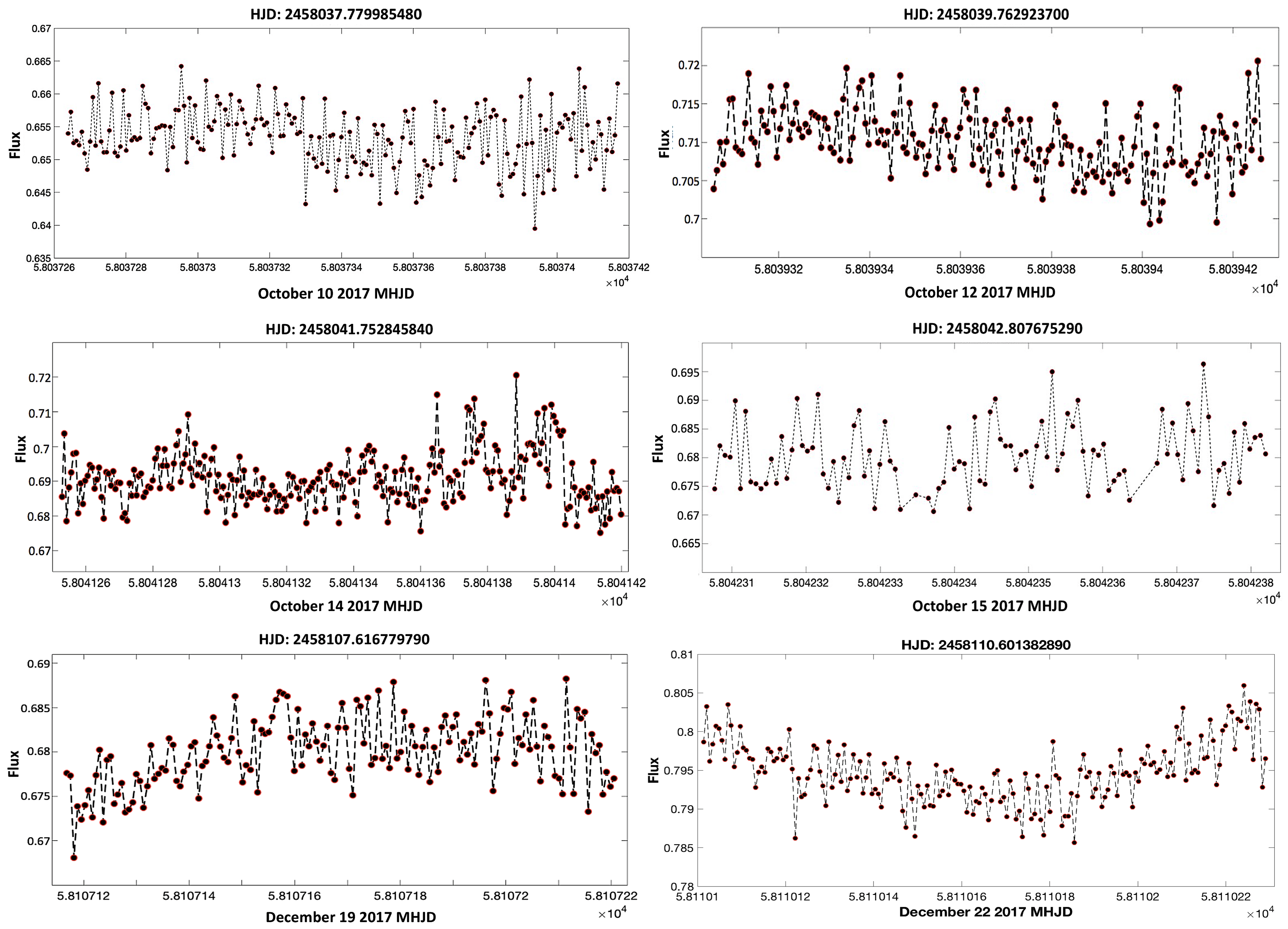}
		\caption{Differential light curves with original (non-normalized) fluxes for BRI 0021, which was monitored for 6 nights over two separate epochs. Throughout the observations, the amplitude varied from 0.0031 - 0.0056 mag. MHJD is Modified Heliocentric Julian Date (HJD - 2,400,000.5 days).}
		\label{fig:BRI LC}
	\end{figure*}

	Extensive radio surveys primarily in the GHz bands indicate that $\sim$ 10\% of local substellar dwarfs are radio active and so have the potential to support aurora in their magnetospheres \citep{pineda2017panchromatic}  evident by the detection of electron cyclotron maser emission (ECM). Whilst this may suggest that such magnetic activity is restricted to a smaller sub-group of such objects, the more likely reason is due to  viewing geometry constraints for the radio emission, as ECM processes beam perpendicular to the local magnetic field, biasing against those dwarfs whose axes of rotation are inclined towards our line of sight, limiting more detections. 
	Consequently, making the not unreasonable assumption that the majority of dwarfs with kG fields have dipolar field orientations approximately aligned with their axes of rotation, those dwarfs that are likely to be detected with active aurorae are expected to have axes that are highly inclined with respect to our line of sight, and they should emit regular beamed coherent radio emission synchronised to the dwarf's rotational period. A good example is the case of the nearby field T6.5 dwarf 2MASS J10475385+2124234, whose radio derived period was determined to be a consequence of co-aligned ECM emission \citep{williams2015}, and whose subsequent rotational period as derived from Spitzer demonstrated evidence for the first direct measurement of zonal atmospheric winds around a brown dwarf \citep{allers2020measurement}, and established a template for studying exoplanetary atmospheres.\\
	
	The inclination angle of rotating stellar object can be determined by accurately measuring both its true rotation period and combining this information with its spectroscopically derived {\it v ~sin ~i} estimate. Extensive spectroscopic surveys to date have provided excellent measurements of the latter for the local substellar dwarf population. The difficulty has been in securing sufficiently sensitive high-speed photometric data over long observational baselines to ensure recovery of the very low duty cycle modulations and so derive unambiguous rotational estimates for the same population, and in particular for that subset of substellar objects for which circumstantial or indeed direct evidence exists for significant magnetic activity, such as a high {\it v sin i}, and the presence of  H$_{\alpha}$ and/or radio emission.\\
	
	Here, we present the findings from $\it I$-band optical monitoring of the nearby rapidly-rotating ultracool M9.5 dwarf BRI 0021-0214, known to exhibit both H$_{\alpha}$and radio emission, carried out using the Galway Ultra Fast Imager (GUFI) photometer with the 1.83 m Vatican Advanced Technology Telescope VATT on Mt. Graham, Arizona. Our goal was to determine evidence for stable periodic optical variability,  likely consistent with that expected from co-rotating magnetic/auroral associated structures in the dwarf's atmosphere, and to use this information to determine its inclination geometry by resolving its known {\it v ~sin ~i}  measure. Relevant information on our target is presented in Section 2, and our observations and data analysis are described in Section 3. In Section 4, we discuss the statistical tools used to evaluate the detected periodic variability. Uncertainties are considered in Section 5. These photometric results, the inclination angle derived from them, and the implications for understanding BRI 0021+0214 are discussed in Section 5.  Our conclusions are summarized in Section 7.
	
	\begin{table*} 
		\centering
		\begin{tabular}[c]{cccccccc}
			
			\hline
			Date of Obs.  & Length of Obs. & Exp. Time & Readout Rate &Amplifier & Reference Stars  \\
			(UT)&  (hr) &(s) &(MHz) &&(No.) \\
			\hline
			2017 Oct 10 &  3.4 & 5 & 1 & Conventional & 1 \\
			2017 Oct 12 &  3.9 & 10 & 1 & Conventional & 1 \\
			2017 Oct 14 &  4 & 5 & 1 & Conventional & 1 \\
			2017 Oct 15 &  2.1 & 5 & 1 & Conventional & 1 \\
			2017 Dec 19 &  2.5 & 5 & 1 & Conventional & 1 \\
			2017 Dec 22 &  3.1 & 5 & 1 & Conventional & 1 \\
			\hline
			
		\end{tabular}
		\caption{\label{tab:BRI OBS}Observation details for BRI 0021.}
	\end{table*}

	\section{BRI 0021-0214}
	
	The M9.5 dwarf BRI 0021-0214 (hereafter BRI 0021) is a rapid rotator ({\it v ~sin ~i}  = $34.2 \pm 1.6$ \text{km} $\text{s}^{-1}$ )  \citep{basri1995surprise}, and is located at a distance of $\sim$11.55 pc \citep{basri1996rotation, reid1999bri}. According to the evolutionary models of \citet{chabrier2000evolutionary}, the absence of lithium in the object's spectrum implies BRI 0021 is no younger than 300 Myr nor less massive than 0.06 M$_{\odot}$. Recently, \citet{gonzales2018understanding} estimated the age of BRI 0021 to be in the range of 500 \text{-} 10000 Myr. \citet{filippazzo2015fundamental} used evolutionary model isochrones to derive mass and radius estimates of $0.075 \pm 0.010$ M$_{\odot}$ and  $0.109  \pm 0.005$ R$_{\odot}$, respectively. Based on spectroscopic observations \citep{leggett2001infrared}, the effective temperature $\it T_{eff}$ and bolometric magnitude (log $\it L_{bol}$/ $\it L_{\odot}$) obtained for this dwarf were, respectively, about 2100K and $\approx$ \text{-} 3.43, consistent with the more recently derived $\it T_{eff}$ and bolometric  magnitude of (log $\it L_{bol}$/ $\it L_{\odot}$) $2385 \pm 77$K and \text{-} $3.44 \pm 0.04$, respectively \citep{gonzales2018understanding}. (see Table~\ref{tab:BRI properties} for the dwarf's physical properties).\\

	\begin{figure*}
		\centering
		%\resizebox{{\columnwidth}{!}{
		\includegraphics[width=15cm]{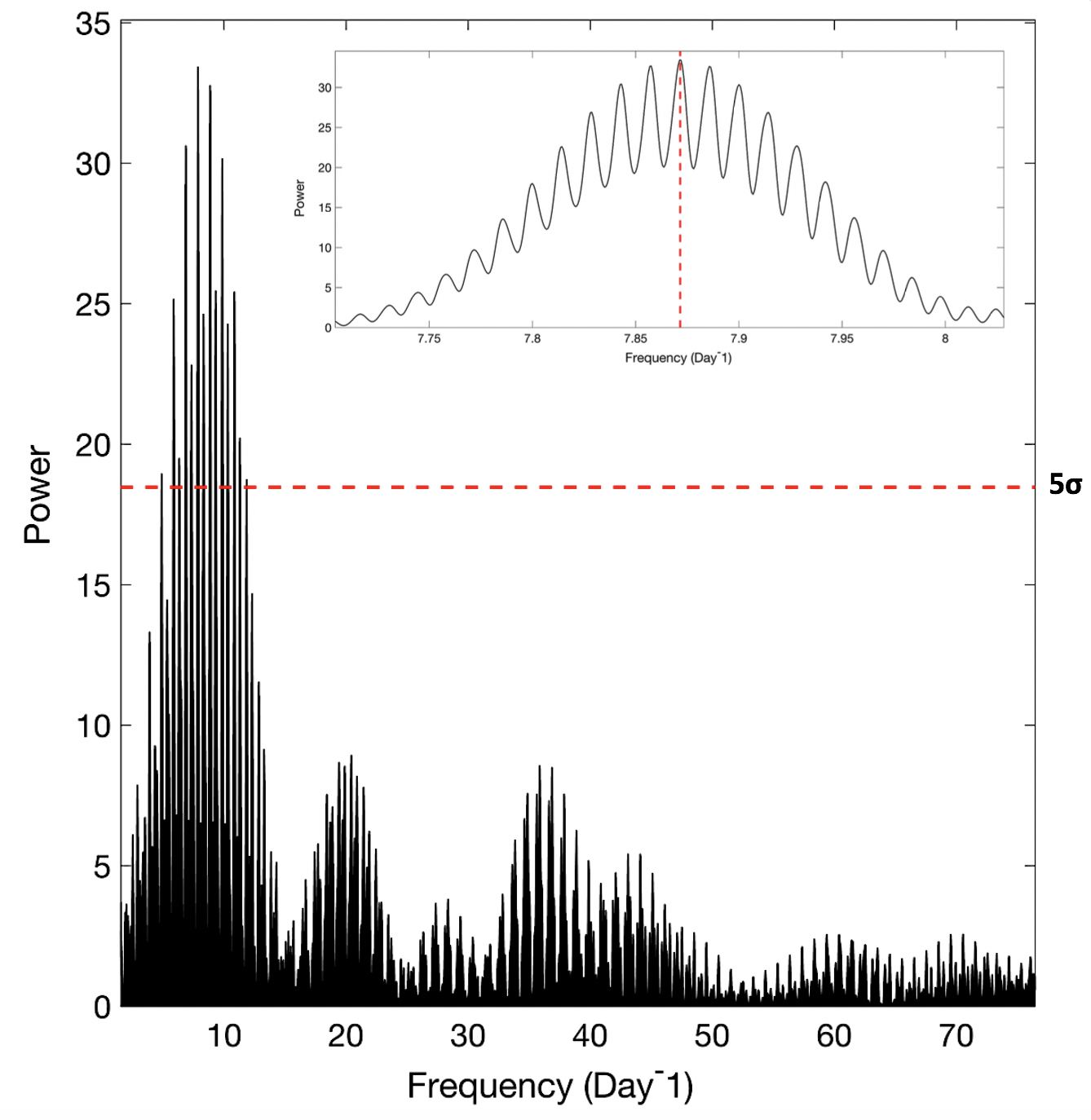}
		\caption{Lomb-Scargle periodogram of all BRI 0021 epochs, calculated from the combined data set in Figure ~\ref{fig:BRI LC}. The red dashed and dotted horizontal line represents a 5 $\sigma$ false-alarm probability of the peaks, as determined by the LS periodogram algorithm. The x-axis is plotted in day$^{-1}$ because each light-curve was time-stamped in units of Heliocentric Julian Date (HJD). The vertical red dashed line at the most significant peak (shown in close-up in the inset plot) corresponds to a period of $3.052 \pm 0.006$ hrs at a significance of $>$ 5 $\sigma$.} 
		\label{fig:BRI LS}
	\end{figure*}
	
	Early observations to discern magnetic activity were not promising. H$_{\alpha}$ emission was not detected \citep{basri1995surprise} nor were observations using ROSAT successful \citep{neuhaeuser1998search}. Subsequent more sensitive observations \citep{reid1999bri} confirmed the presence of a magnetic field in this object by detecting extremely low levels of H$_{\alpha}$ flare emission with a luminosity log  $\it L_{\alpha}$ / $\it L_{bol}$ $\approx$ $10^{-4.2}$, a factor of 3 lower than the average level of quiescent emission of early- to mid-type M dwarfs, and this was followed by the detection of low-level radio emission (average value of $40 \pm 13$ $\mu$Jy at 8.46 GHz) with peak flux densities of $360 \pm 70$ $\mu$Jy from the dwarf \citep{berger2002flaring}. Subsequent simultaneous X-ray, radio and H$_{\alpha}$ observations demonstrated clear evidence for a steady variability of H$_{\alpha}$ emission from BRI 0021 on a timescale of 0.5 \text{-} 2 hours, although there were no detected X-ray or radio emissions \citep{berger2009simultaneous}.\\

	A photometric campaign was carried out by \citet{martin2001photometric} to search for the presence of optical variability. They found $\it I$-band variability with periods of $\sim$20 hrs and $\sim$4.8 hours. They inferred that the variability was likely not due to spots on the stellar surface, since preliminary studies at that time showed the dwarf appeared to have low levels of magnetic activity. Rather, they argued that the cause would be due to dust clouds in the dwarf's atmosphere, as there is evidence of silicate and iron clouds in the spectrum of BRI 0021 \citep{chabrier2000evolutionary,ackerman2001precipitating}. However when \citet{miles2013linear} studied near-infrared linear polarization of rapidly rotating ultracool dwarfs from late-M to early-T dwarfs, they detected no linear polarization of BRI 0021 at optical or near-infrared wavelengths during different observation epochs, which might have been expected for a dust-dominated modulated signal \citep{osorio2005optical}.\\

	We previously \citep{harding2013periodic} observed BRI 0021  for a total of $\sim$ 28 hours using the same instrumentation configuration described in this paper (see Section 3), with our analysis presenting evidence for aperiodic variations, possible periodic variability with a period between 4 \text{-} 7 hours, with the strongest evidence for a periodic solution of  $\sim$ 5 hours albeit at low significance. Although in agreement with the faster $\sim$ 4.8 hours period found by \citet{martin2001photometric},  neither measured period  agrees with a maximum period of about 3.59 hrs inferred from {\it v ~sin ~i} observations \citep{mohanty2003rotation}. The observations reported by \citet{harding2013periodic} consisted of three separate observational epochs, the first in September 2009, and the remaining two in November and December 2010, with the latter separated by 17 nights. In each case, 9 hours of data were taken over two consecutive nights (for the 2010 data) and two nights separated by one unobserved night (for 2009). The lack of any consistency in the results reported by \citet{harding2013periodic} is not entirely unexpected, given BRI 0021's known variability across multiple passbands and in particular the timecales associated with these latter observations. With that in mind, we decided to implement a follow-up observation campaign for this object using GUFI \textit{VATT}over a  shorter observational baseline to definitively estimate the presence, if any, of any unambiguous rotational signature for this dwarf consistent with sources in the atmosphere/photosphere with limited intrinsic duty cycles.

	\begin{figure*}
		\centering
		%\resizebox{{\columnwidth}{!}{
		\includegraphics[width=15cm]{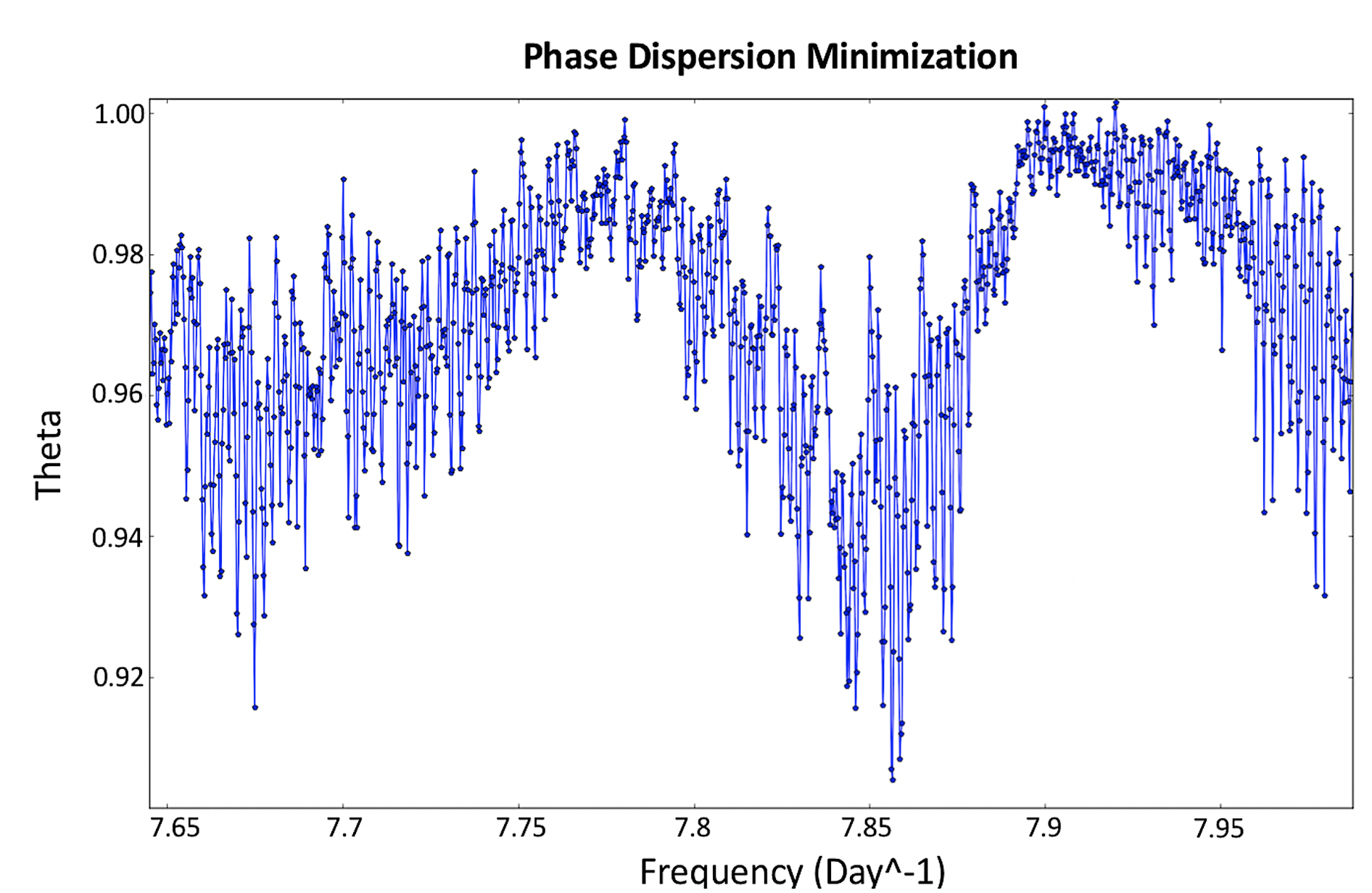}
		\caption{Phase dispersion minimization plot of all of the BRI 0021 data \text{-} Theta statistic against frequency ($\text{Day}^{-1}$) for $10^4$ Monte-Carlo simulations. The minimum theta corresponds to $\sim$ 3.05 hrs, which is largely consistent with the rotation period resulting from the LS periodogram.}
		\label{fig:BRI PDM}
	\end{figure*}

	\section{Photometric Observations and Analysis}
	
	We obtained observations of BRI 0021 on four nights in October 2017 (${10^\text{th}}$, ${12^\text{th}}$, ${14^\text{th}}$ and ${15^\text{th}}$) and on two nights in December 2017 (${19^\text{th}}$ and ${22^\text{nd}}$) for a total of $\sim$19 hours, as shown in Table ~\ref{tab:BRI OBS}. These were carried out with the GUFI photometer on the 1.83m VATT, Mt. Graham International Observatory, Arizona. Integrated with the VATT, GUFI offers a FOV of 3$'$$\times$3$'$ and a plate scale of 0.35$''$ pixel$^{-1}$, and its L3-CCD can operate at up to 34 frames per second in full-frame readout, with frame transfer times $\sim$ 2 ms  \citep{harding2013periodic}, so "dead time" is negligible. The data were acquired through the VATT {\it I}-Arizona ($\sim$ 7200 - 9100 \AA) broadband filter, with exposure times of 5 and 10 sec (see Table ~\ref{tab:BRI OBS}), and seeing between 0.9$''$ to 1.7$''$. The frames were later summed into 1-minute images, to increase the signal to noise (SNR) while preserving time resolution appropriate to the target.\\
	
	The data was reduced using our in-house GUFI pipeline \citep{Sheehan2008, sheehan2008development}, which implemented standard techniques of data reduction on all frames. The science images were de-biased using nightly masterbias frames, and then flat-fielded using twilight flat-fields, which for any given observation, were made up of more than 90 median-combined dithered frames taken from a blank part of the sky. A de-fringing routine was then applied, using a fringing template produced from the median of a large number of dithered, reduced science frames. The frames were then registered to correct any potential guiding drift, and co-added in 1-minute batches to increase the SNR. 
	The selected reference star: VHS J002423.70-015906.0 $(00^h 24^m 23^s.70648 -01^{\circ}$ 59$'$ 06$''$.0648) used is the same as the comparison star selected by \citet{martin2001photometric} and later by \citet{harding2013periodic} based on its observed stability relative to the target star.
	The pipeline carried out differential photometry on the reduced images, measuring both the target BRI 0021 and a selected reference star, in order to ensure that the variability was indeed inherent to the target. Only one reference star meeting the criteria in \citet{bailer2001variability} was available within the GUFI FOV while observing BRI 0021. The differential light curves were evaluated to check for any remaining systematic trends due, for example, to the effects of increasing/decreasing air mass. These effects were virtually nil throughout all the observations, where conditions were photometric.  \\

	\begin{figure*}
		\centering
		%\resizebox{{\columnwidth}{!}{
		\includegraphics[width=16cm]{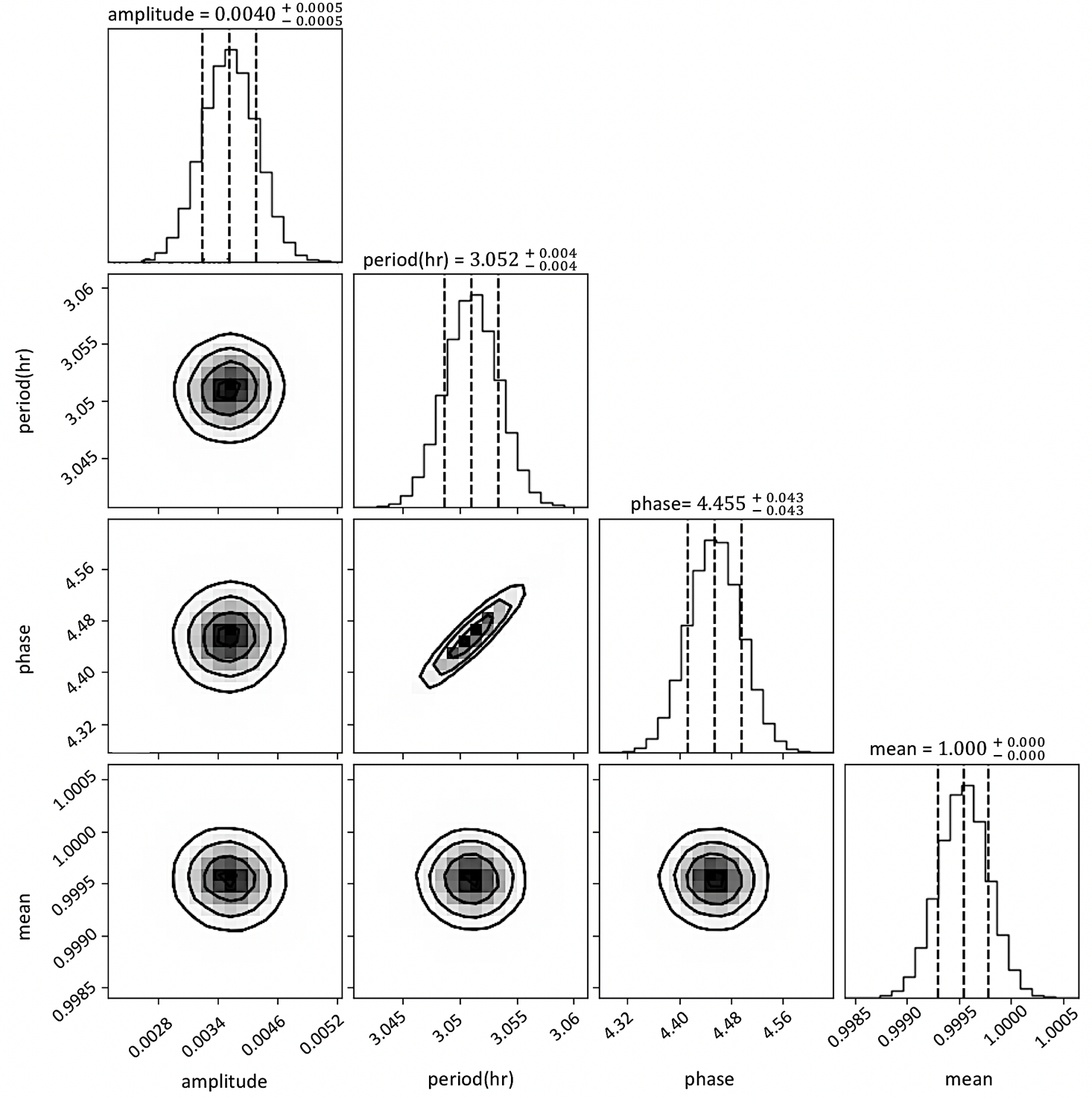}
		\caption{Posterior probability distributions obtained from applying the MCMC algorithm, in this case, for BRI 0021 time series data obtained on October 14 2017. Vertical dashed lines show the median and 1 $\sigma$ uncertainties on the best-fitting model parameters, which are defined by the 16 \% and 84 \% quantiles of each distribution. Contours show the 0.5, 1.0, 1.5, and 2.0  $\sigma$ uncertainties of the posterior probabilities for each two-dimensional parameter.).}
		\label{fig:Figure4_MCMC_BRI_14_10_2017}
	\end{figure*}

	%\begin{landscape}
	
	%\end{landscape}
	
	%\begin{landscape}
	% \begin{figure*}
	%\centering
	%\resizebox{{\columnwidth}{!}{
	%\includegraphics[width=9cm]{Figure5_NONNormalized}
	%\caption{These plots show the measured non-normalized to unity differential BRI 0021 time series for the nights of October 10 2017, October 14 2017 and 22 Dec 2017. We note the 'jump' in diff flux for the lower panel of Figure 2, due to the night Dec 22 is slightly larger in amplitude than the other nights (see Figure ~\ref{fig:BRI LC} and Table ~\ref{tab:BRI OBS}). }  
	%\label{fig:Figure5_NONNormalized}
	%\end{figure*}
	%\end{landscape}

	\section{Periodic Variability Analysis}
	
	Having extracted the differential photometric time-series of BRI 0021 for each run, we used Lomb-Scargle (LS) periodogram
	analysis \citep{lomb1976least, scargle1982studies, scargle1989studies}, phase dispersion minimization (PDM) \citep{stellingwerf1978period} and  a Markov Chain Monte Carlo (MCMC)  sinusoid model fitting method \citep{allers2020measurement}, to determine the presence of any underlying periodic signals. For the former we chose a range of LS power peaks above the noise-floor ( $>$ 5 $\sigma$ significance), and for the PDM analyses, we performed $10 ^ 5$ Monte-Carlo simulations per assay in order to evaluate the statistical significance of the potential outcomes. For the MCMC estimation of sinusoidal variability, we used 500 walkers with 20000 steps, following a burn-in sample of 1000 steps for each case. 
	The data was then phase-folded based around the most significant periodicities from these analyses. We assessed the amplitude variability of each target light curve via a sinusoidal fitting model using the identified rotation period, and varied the amplitude and phase of the sinusoidal function. A ${\chi}^2 $ goodness-of-fit test was performed to give the best fit for a given night's amplitude variability, which can also estimate the error of the amplitude variability, period and phase of a given photometric light curve.\\

	\begin{figure*}
		\centering
		%\resizebox{{\columnwidth}{!}{
		\includegraphics[width=15cm]{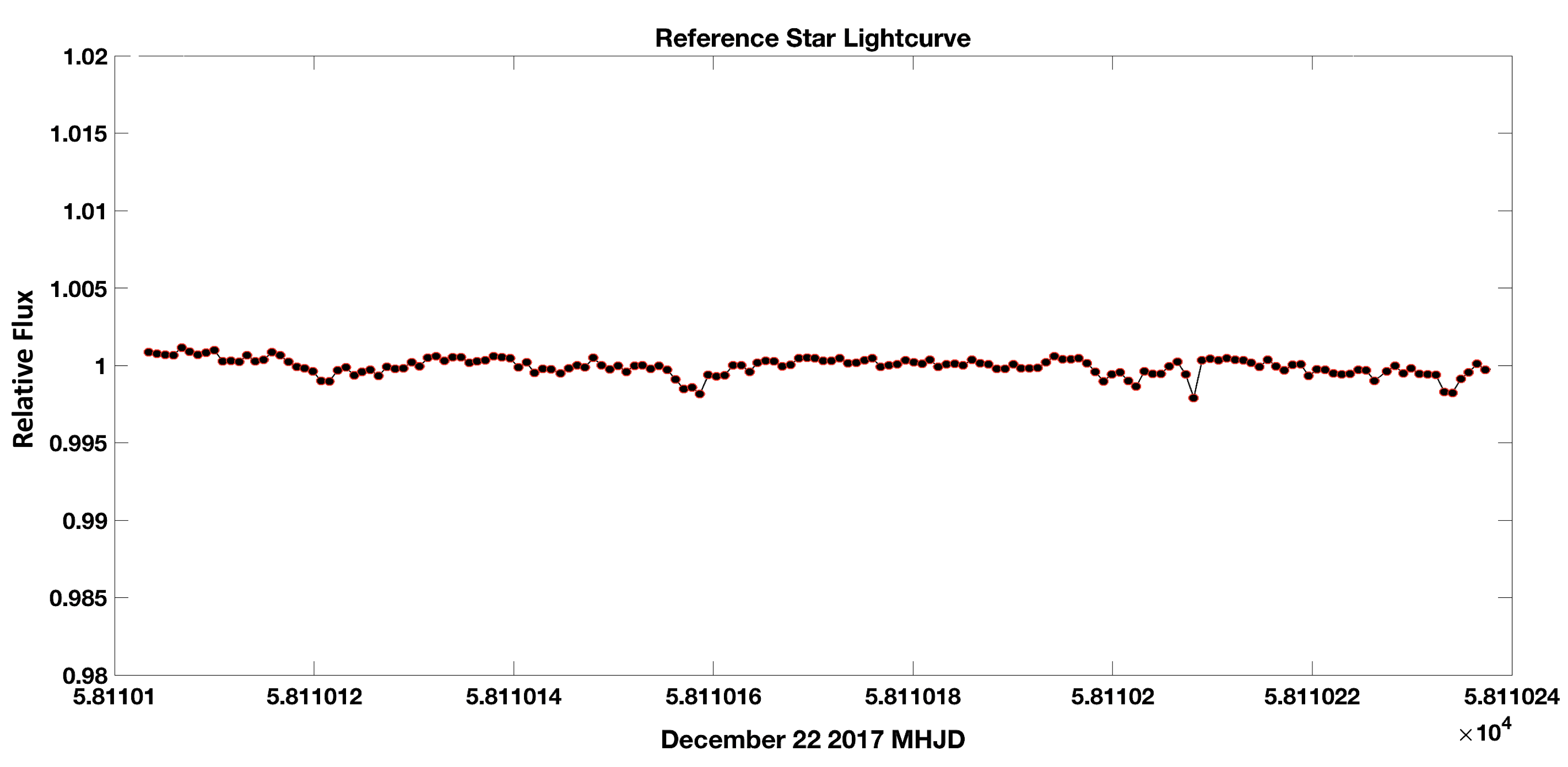}
		\caption{We show VHS star light curve for the night 22 Dec 2017, to illustrate the non-variability of the comparison star used in the differential photometry of BRI 0021-0214.}
		\label{fig:refrence_22_Dec}
	\end{figure*}

	\section{Estimation of the Period Uncertainty and Photometric Error}
	
	Magnitude errors in the target and reference star data points were computed via the IRAF task {\tt \small phot}, and then converted to flux errors to match the light curve units. The photometric error model described by \citet{bailer2001variability} was used to quantify formal photometric errors (resulting from photon statistics and instrumental readout noise) and informal photometric errors (such as residuals after flat fielding and fringe removal). \\ 
	
	To estimate the period uncertainty for the potential rotation periods detected for our target, we overplotted the LS power spectrum period range with a Gaussian profile, and then measured the FWHM of the strongest peak, since the width of the FWHM covered a range of possible true periods that could lie within it.The period uncertainty ($\delta P$) is calculated as follows:
	
	\begin{equation}
	\centering
	\label{eq : period uncertainty}
	\delta P = \frac{\text{FWHM}}{{2.35482}}
	\end{equation}
	
	where FWHM=$\sqrt{2 ln 2}$.$\sigma$ = 2.35482$\sigma$. This gives an estimate of the period uncertainty to 1$\sigma$. We used this technique with each night's light curve to calculate the uncertainty range for the detected rotation period. Using the Markov Chain Monte Carlo (MCMC) algorithm provides period uncertainty via the derived posterior probability distribution.

	\begin{table*} 
		\centering
		\begin{tabular}[c]{ccccccccc}
			
			\hline
			Date of Obs. & Band & $ \text{Amplitude Variability}_\text{Target}$  & Photometric Error\\
			(UT)&  & (mag) &(mag) \\
			(1)&(2)&(3)&(4)\\
			\hline
			2017 Oct 10& $\it I$ & $0.0048 \pm 0.0003$  &0.0046\\
			2017 Oct 12 & $\it I$ & $0.0052 \pm 0.0003$  &0.0028\\       
			2017 Oct 14 & $\it I$ &$0.0040 \pm 0.0004$& 0.0031\\
			2017 Oct 15 & $\it I$ &$0.0038 \pm 0.0003$& 0.0047\\
			2017 Dec 19 & $\it I$ &   $0.0031 \pm 0.0004$&0.0058\\
			2017 Dec 22 & $\it I$ & $0.0056\pm 0.0004$& 0.0073\\          
			\hline
			
		\end{tabular}
		\caption{\label{tab:BRI Var}Amplitude variability and photometric error analysis of BRI 0021}.{ 
			\textbf{Column 1}: Date of observation in UT. \textbf{Column 2}: Waveband used. \textbf{Column 3}: Peak-to-peak amplitude variability with associated errors as measured by the ${\chi}^2 $ test. \textbf{Column 4}: Nightly mean photometric error bar for light curve data points, as measured by the \emph{iraf.phot} routines.}
	\end{table*}

	\section{Results}
	We carried out fixed aperture differential photometry using the L3-GUFI pipeline and found {\it I}-band periodic variability with mean amplitude variability of 0.0031 to 0.0056 mag, which is shown in Figure ~\ref{fig:BRI LC} \& Supplementary Figure 1 (normalised to unity). The 3$'$$\times$3$'$ field of view (FOV) of GUFI is relatively small and in this case, it only contained one suitable comparison star, which has been used by all previous authors studying optical variability for BRI 0021. \citet{martin2001photometric} identified two periods for BRI 0021 which were $\sim$ 4.8 hrs and $\sim$ 20 hrs, respectively. However, \citet{harding2013periodic} showed possible solutions of $\sim$ 4 to 7 hours. Because we observed BRI 0021 for a limited time of $\sim$ 19 hours, we do not have enough time coverage to evaluate the hypothesised 20 hours period. However, our analysis of the differential magnitudes, timestamped with Heliocentric Julian Date, revealed sinusoidal periodic variability of $3.052 \pm 0.006$ hours (as shown in Figures ~\ref{fig:BRI LS}, ~\ref{fig:BRI PDM}, Supplementary Figure 2, Supplementary Figure 3 \& Table ~\ref{tab:BRI Var}). In Figure ~\ref{fig:Figure4_MCMC_BRI_14_10_2017} we show the results of an example MCMC analysis for 14 October 2017, illustrating the performance of this inference algorithm in recovering a periodic signal incorporating uncertainties on the parameters via the derived posterior probability distribution. 
	The MCMC method gives a rotational period of $3.052 \pm 0.004$ hours.\\
	
	We investigated the comparison star: VHS J002423.70-015906.0 $(00^h 24^m 23^s.70648 -01^{\circ}$ 59$'$ 06$''$.0648) to examine if the photometric variability was caused by the reference source, BRI 0021, or both (see Figure ~\ref{fig:refrence_22_Dec}). As a result, we can confidently state that the short period of $3.052$ hours is due to BRI 0021, which is known to be a fast rotator.\\
	
	\begin{figure*}
		\centering
		%\resizebox{{\columnwidth}{!}{
		\includegraphics[width=15cm]{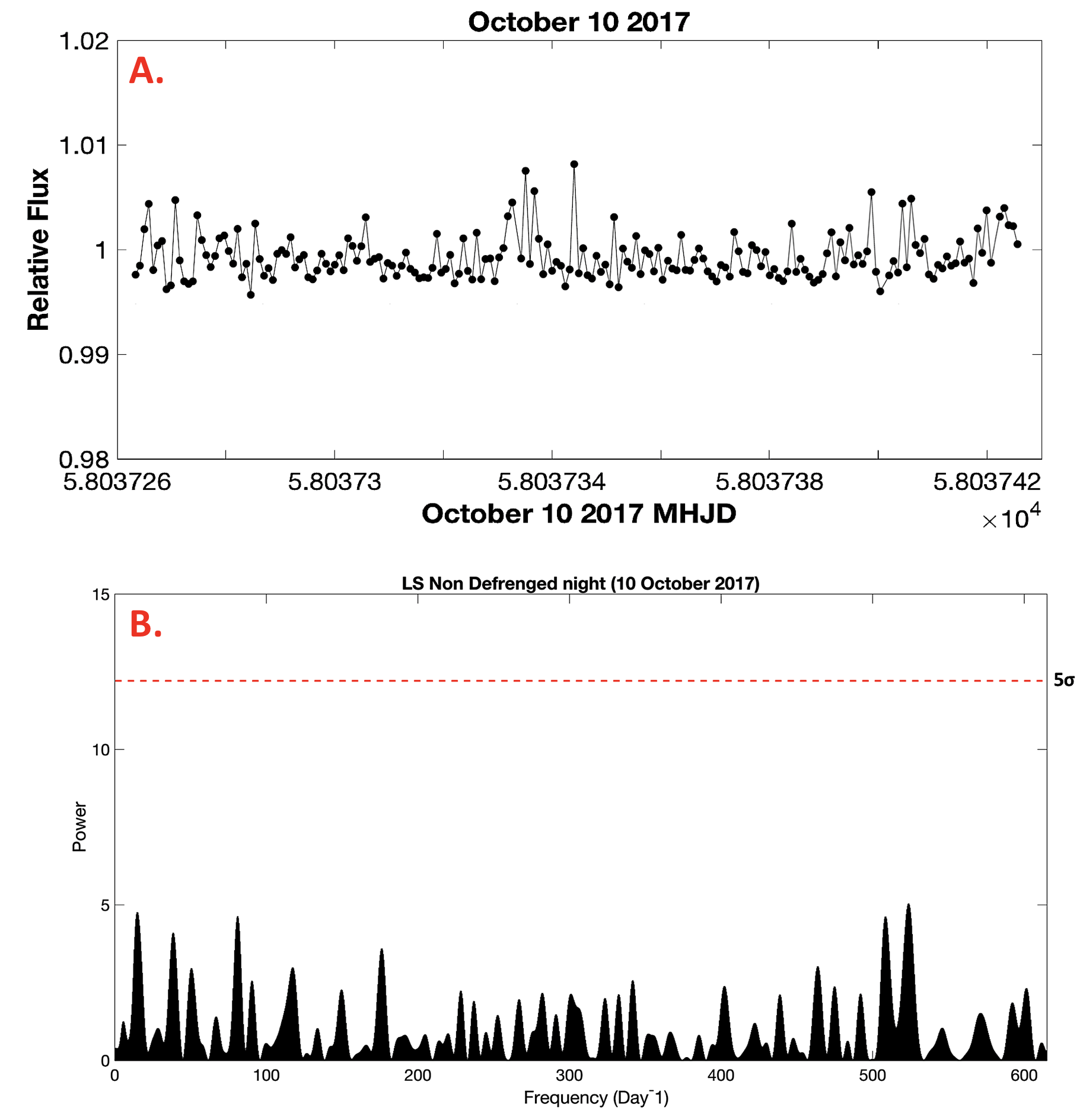}
		\caption{{\color{red} A.} Time-series photometry of the non-defringed frames (for the night of October 10 2017 in this case) on a sky background area to determine if the fringe amplitude has a temporal signature. {\color{red} B.} Lomb-Scargle periodogram of the same time-series, confirming a lack of fringe-associated variability.}
		\label{fig:Figure6_NON_defrenged}
	\end{figure*}
	
	For all nights the resulting periods and uncertainties were in agreement using this approach, with an overall best estimate for the rotational period of $3.052 \pm 0.004$ hours. Our period estimates are in consistent with with the rotational velocity of $34 \pm 2$ \text{km} $\text{s}^{-1}$ found by \citet{mohanty2003rotation}, which implies a maximum possible rotation period for this system of $\sim$ 3.59 hours.  \\
	
	%{\color{red} In figure ~\ref{fig:Figure4_MCMC_BRI_14_10_2017}, we also used Markov Chain Monte Carlo (MCMC) algorithm to determine and evaluate the period of this object which will naturally include uncertainties on the parameters via the posterior probability distribution, as we selected randomly on the night of 14 October 2017. The resulting periods and uncertainties of all statistical analysis are in agreement with each other. The MCMC method gives a rotational period of $3.052 \pm 0.004$ hours.}\\
	
\begin{figure*}
		\centering
		%\resizebox{{\columnwidth}{!}{
		\includegraphics[width=15cm]{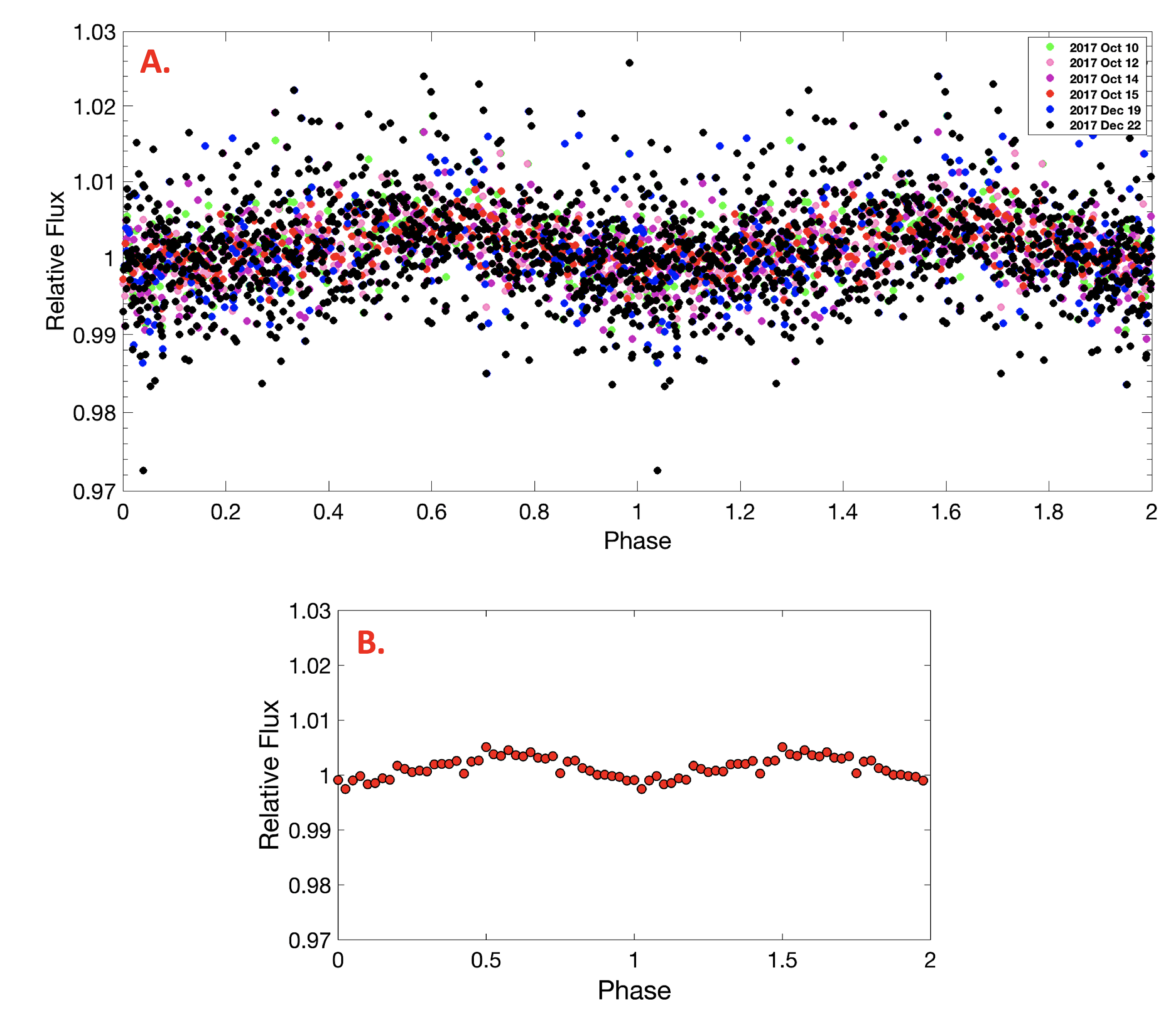}
		\caption{{\color{red} A.} Phase-folded light curve of all epochs in the BRI 0021 observation campaign. {\color{red} B.} The same, except binned by a factor of 10. Each light curve was phase-folded on the detected period of $3.052 \pm 0.004$ hrs.}
		\label{fig:BRI_binned}
		\label{{fig:BRI_binned}}
	\end{figure*}

	We note that each of the six observations had temporal baselines similar to our estimated rotational period for BRI 0021, however, any sampling effects resulting in a systematic periodic signal at this approximate frequency, and any associated harmonics, would be sychronised to all sources and so removed during the differential step. However, we recovered the same periodic signal from individual nights (e.g. Figure \ref{fig:Figure4_MCMC_BRI_14_10_2017}) for which no such sampling bias or composite window function could have been present. 
	We performed timeseries photometry on the non-defringed frames, rather than the BRI 0021 target or reference stars, to determine if there is a temporal signature in the fringe amplitudes. The light curves all appeared to be consistent with random noise. Performing Lomb-Scargle (LS) periodograms analysis confirmed the lack of any detectable periodic behaviour due to fringe amplitude variations (Figure \ref{fig:Figure6_NON_defrenged}). Taken together these further substantiate the detection of a periodic signature from BRI 0021 in this work -  in Figure ~\ref{fig:BRI_binned} we applied a normalization to the mean fluxes per night for producing the phase folded light curve on $3.052 \pm 0.004$ hours.\\

\begin{figure*}
		\centering
		%\resizebox{{\columnwidth}{!}{
		\includegraphics[width=18cm]{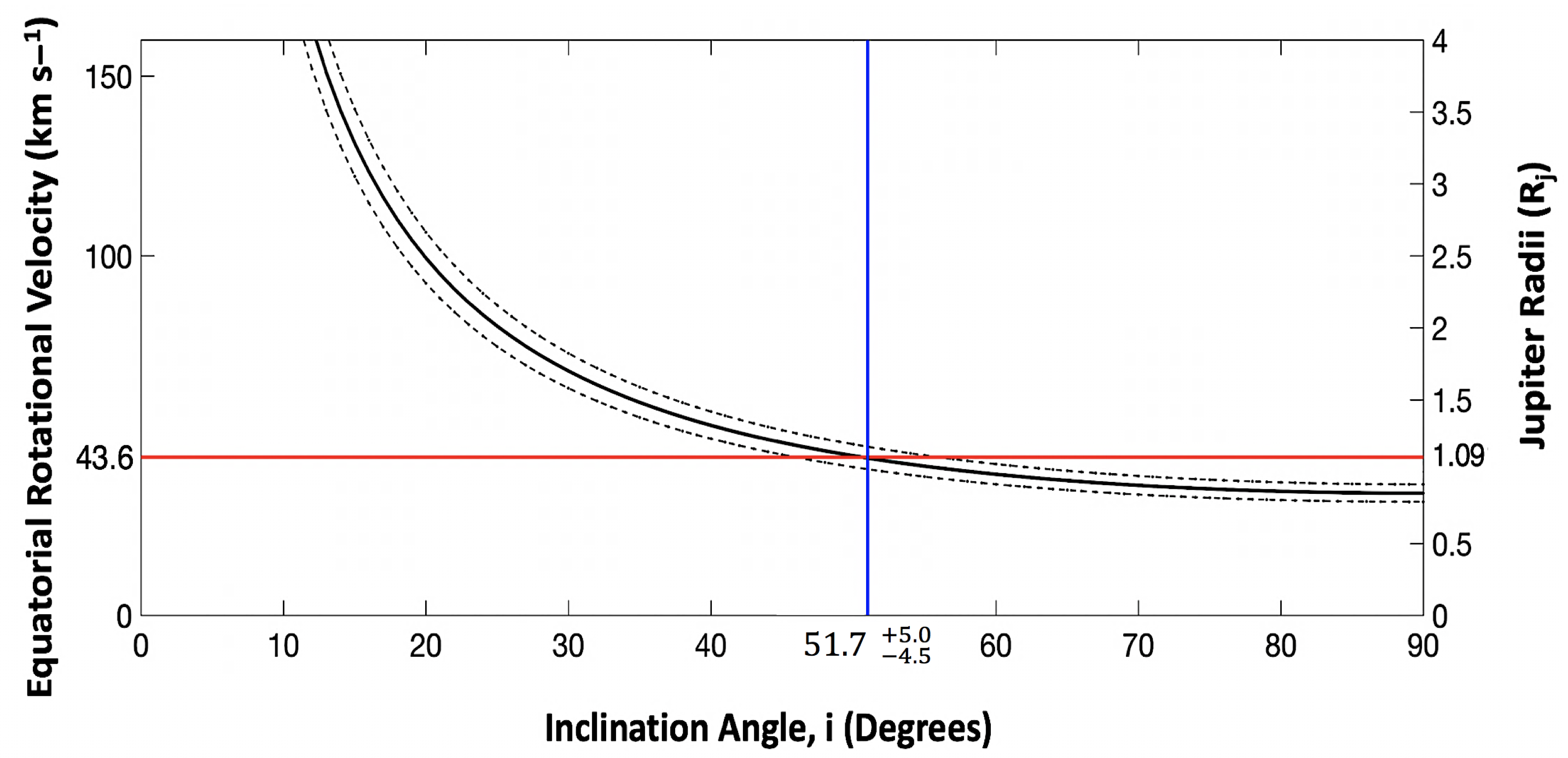}
		\caption{The {\it v sin i}  of BRI 0021, $34.2 \pm 1.6$ \text{km} $\text{s}^{-1}$ \citep{crossfield2014doppler}, is shown by the solid black curve, with dashed lines representing the associated errors. It constrains the equatorial rotational velocity vs. the inclination angle for the rotation axis of BRI 0021 relative to our line of sight. The solid blue vertical line highlights the inclination angle. The second Y-axis, right, is the radius of the dwarf, tracking the equatorial rotational velocity. The predicted radius of $1.09 \pm 0.05$  $\text{R}_\text{J}$ from \citet{filippazzo2015fundamental} is marked by a solid red line where  $\text{R}_\text{J}$ $\sim$ 69911 km, as calculated in this work, and indicates an equatorial velocity of $43.6 \pm 2$ \text{km} $\text{s}^{-1}$.  }
		\label{fig:BRI vsini}
	\end{figure*}

	Assuming our measured periodic signal corresponds to the rotational period of BRI 0021, this data allows us to break the {\it v sin i}  degeneracy and so calculate the inclination angle for the rotation axis of BRI 0021 relative to our line of sight. Using the dwarf's derived radius of $ 0.109 \pm 0.005 $ R$_{\odot}$ \citep{filippazzo2015fundamental}, the most recent  {\it v ~sin ~i}  of $34.2 \pm 1.6$ km $\text{s}^{-1}$ \citep{crossfield2014doppler}, and our estimated rotation period, we derive an inclination angle of 51.7$^{+5.0}_{-4.5}$ degrees for the rotation axis relative to our line of sight, and an equatorial rotational velocity of $43.6 \pm 2$ km $\text{s}^{-1}$, as shown in Figure ~\ref{fig:BRI vsini}. The derived radius of 0.109 R$_{\odot}$ from \citet{filippazzo2015fundamental} is in good agreement with the predicted radius of $\sim$  0.11 R$_{\odot}$ from evolutionary models \citep{chabrier2000evolutionary}, and consistent with the model predictions in \citet{martin2001photometric} where the radius of BRI 0021 should be $\leq$ 0.11 R$_{\odot}$. \\
	
	We sought to determine if this rotational estimate could bring any clarity to the previous observations of BRI 0021. We took each of the three previous observational epochs of \citet{harding2013periodic} and folded each photometric time series using a period of 3.052 hours. We combined all three datasets and repeated this process. In all cases, no consistent modulated signal was apparent, nor did any resulting lightcurves have any significance using standard goodness-of-fit tests.
	We subsequently combined our 2017 epoch data with those of \citet{harding2013periodic}, and on implementing our periodicity analysis noted no change in the derived period - albeit with an increased error estimate - demonstrating that the 2009/2010 observational data provide no additional information. Randomly shuffling the \citet{harding2013periodic} data and repeating this combined periodicity analysis yielded a similar outcome. Taken together, the most reasonable conclusion to draw from this is that the observations made by \citet{harding2013periodic} correspond to times when the surface conditions of BRI 0021-0214 were consistent with those observed by other authors; that is to say, no evident stable structures - whether these be localised meteorological phenomena, magnetic or auroral spots - were in existence at these earlier epochs. \\

	As stated in Section 2, BRI 0021 has been associated with intermittent H$_{\alpha}$ flaring activity, and coherent radio emission, certainly suggestive of magnetic activity on this M9.5 dwarf. Under the auroral model, best demonstrated to be operating within the magnetosphere of the nearby M8.5 dwarf LSR J1835+3259, beamed radio emission is coincident with an auroral oval emitting in H$_{\alpha}$  and the underlying atmosphere heated by associated particle beams yielding differing radiated response as a function of frequency depending on the resulting plasma's optical thickness \citep{hallinan2015magnetospherically}. Critical to the radio beams' visibility is the inclination angle of the dwarf's rotation axes to our line of sight, as when coherent radio emission originates from the electron cyclotron maser instability, the maximum intensity is beamed perpendicular to the dwarf's magnetic dipolar rotation axis, believed to be the source of all observed GHz radio emission observed to date from the sub-group with functioning aurora. If we make the not unreasonable assumption that such dipolar fields are expected to be co-aligned with the dwarf's rotational axis, then for the case of BRI 0021, with an inclination angle of 51.7$^{+5.0}_{-4.5}$ degrees, significantly less beamed emission is anticipated, although any particle beams impacting the dwarf's atmosphere would be likely to yield a discernible modulated optical signature. Given the intermittent nature of BRI 0021's H$_{\alpha}$ and radio emission, combined with the weak optical signal presented in this work, the evidence is  not inconsistent with an intermittent auroral mechanism associated with it's magnetosphere. That being said, this intriguing dwarf's variability is likely due to more than one astrophysical cause. Time-evolving atmospheric
	dust clouds could periodically conceal the cool/hot spots on the stellar surface \citep{lane2007rotational,heinze2013weather,metchev2015weather}, with likely intermittent auroral activity previously described playing an
	occasional role in this process \citep{hallinan2015magnetospherically}.\\

	\section{Conclusions}
	We report the results of a recent $\it I$-band photometric study of the nearby M9.5 field dwarf BRI 0021 using the GUFI high speed photometer. This object's detected periodic variability and associated period were assessed using two analysis methods. However, period errors were calculated by fitting a Gaussian function to the highest Lomb-Scargle periodogram peak. In fact, this can overestimate the uncertainties on the derived periods \citep{vanderplas2018understanding}. Consequently we adopt MCMC derived errors.\\ 
	
	Over a 73 day baseline, we determined a consistent periodic signal of $3.052 \pm 0.004$ hours, well within the maximum bound of $\sim$ 3.59 hrs inferred from previous {\it v sin i} observations \citep{mohanty2003rotation}. Taking this value to be the dwarf's actual rotational period allows us to infer it's inclination angle with respect to our line of sight of 51.7$^{+5.0}_{-4.5}$ degrees. Such a viewing angle is likely to significantly diminish the anticipated duty cycle associated with the presence of any suspended dust clouds and/or auroral ovals co-rotating with the dwarf, and in particular, to significantly diminish the likelihood of auroral associated beams of coherent radio emission crossing our line of sight; and it is certainly consistent with observations made to date of this rapidly rotating substellar object. Further multiwavelength and spectropolarimatric observations are required to determine if BRI 0021 hosts a functioning auroral mechanism, along with the necessary stable and structured kG fields to sustain such magnetospheric processes, or whether this dwarf is the exception to the rule as regards the subset of known radio-active dwarfs spanning the substellar boundary.

	\section*{Acknowledgements}
	S.D. gratefully acknowledges financial support from the Ministry of higher education in Iraq and Al-Nahrain University who funded this work, and the Dean of Graduate Studies at University of Galway. We also wish to acknowledge the VATT telescope group for their cooperation and guidance. A.G. acknowledges support from an Athena Swan Midcareer Lecturer Capacity Building Grant award from University of Galway's Office of the Vice-President for Equality \& Diversity.  
	
	%%%%%%%%%%%%%%%%%%%%%%%%%%%%%%%%%%%%%%%%%%%%%%%%%%
	
	
	\section*{Data availability}
	The raw GUFI data underlying this article are available in the University of Galway / Centre for Astronomy data repository, and can be accessed at https://tinyurl.com/BRI0021

	\bibliographystyle{mnras}
	\bibliography{BRI_revised_bib} % if your bibtex file is called example.bib

	%\begin{thebibliography}{99}
	%\end{thebibliography}
	
	%%%%%%%%%%%%%%%%%%%%%%%%%%%%%%%%%%%%%%%%%%%%%%%%%%
	
	%%%%%%%%%%%%%%%%% APPENDICES %%%%%%%%%%%%%%%%%%%%%

	%%%%%%%%%%%%%%%%%%%%%%%%%%%%%%%%%%%%%%%%%%%%%%%%%%

	% Don't change these lines
	\bsp	% typesetting comment
	\label{lastpage}
\end{document}